\documentclass[sigconf,nonacm]{acmart}
\usepackage{booktabs}
\usepackage{longtable}
\usepackage{xcolor}
\usepackage{geometry}
\usepackage{tabularray}
\usepackage{tablefootnote}
\usepackage{array}
\usepackage{graphicx}
\usepackage{multirow}
\usepackage{ragged2e}
\usepackage{color}
\usepackage{footnote}

\usepackage{amssymb} 

\AtBeginDocument{%
  }

\copyrightyear{2025}
\acmYear{2025}
\setcopyright{cc}
\setcctype{by}
\acmConference[FAccT '25]{The 2025 ACM Conference on Fairness, Accountability, and Transparency}{June 23--26, 2025}{Athens, Greece}
\acmBooktitle{The 2025 ACM Conference on Fairness, Accountability, and Transparency (FAccT '25), June 23--26, 2025, Athens, Greece}\acmDOI{10.1145/3715275.3732017}
\acmISBN{979-8-4007-1482-5/2025/06}

\begin{document}

\title{Recourse, Repair, Reparation, \& Prevention: A Stakeholder Analysis of AI Supply Chains}


\author{Aspen K. Hopkins}
\affiliation{%
 \institution{Massachussetts Institute of Technology}
 \city{Cambridge}
 \state{Massachussetts}
 \country{USA}
 }
 \email{dataspen@mit.edu}

\author{Isabella Struckman}
\affiliation{%
  \institution{Massachussetts Institute of Technology}
  \city{Cambridge}
  \state{Massachussetts}
  \country{USA}}

\author{Kevin Klyman}
\affiliation{%
  \institution{Stanford}
  \city{Stanford}
  \state{California}
  \country{USA}}

\author{Susan S. Silbey}
\affiliation{%
  \institution{Massachussetts Institute of Technology}
  \city{Cambridge}
  \state{Massachussetts}
  \country{USA}}

\renewcommand{\shortauthors}{Hopkins et al.}

\begin{abstract}
The AI industry is exploding in popularity, with increasing attention to potential harms and unwanted consequences. In the current digital ecosystem, AI deployments are often the product of AI supply chains (AISC): networks of outsourced models, data, and tooling through which multiple entities contribute to AI development and distribution. AI supply chains lack the modularity, redundancies, or conventional supply chain practices that enable identification, isolation, and easy correction of failures, exacerbating the already difficult processes of responding to ML-generated harms. As the stakeholders participating in and impacted by AISCs have scaled and diversified, so too have the risks they face. In this stakeholder analysis of AI supply chains, we consider \textbf{who} participates in AISCs, \textbf{what} harms they face, \textbf{where} sources of harm lie, and \textbf{how} market dynamics and power differentials inform the type and probability of remedies. Because AI supply chains are purposely invented and implemented, they may be designed to account for, rather than ignore, the complexities, consequences, and risks of  deploying AI systems. To enable responsible design and management of AISCs, we offer a typology of responses to AISC-induced harms: recourse, repair, reparation or prevention. We apply this typology to stakeholders participating in a health-care AISC across three stylized markets---vertical integration, horizontal integration, free market---to illustrate how stakeholder positioning and power within an AISC may shape responses to an experienced harm.
\end{abstract}






\maketitle

\begin{figure*}
    \centering
    \includegraphics[width=.95\linewidth]{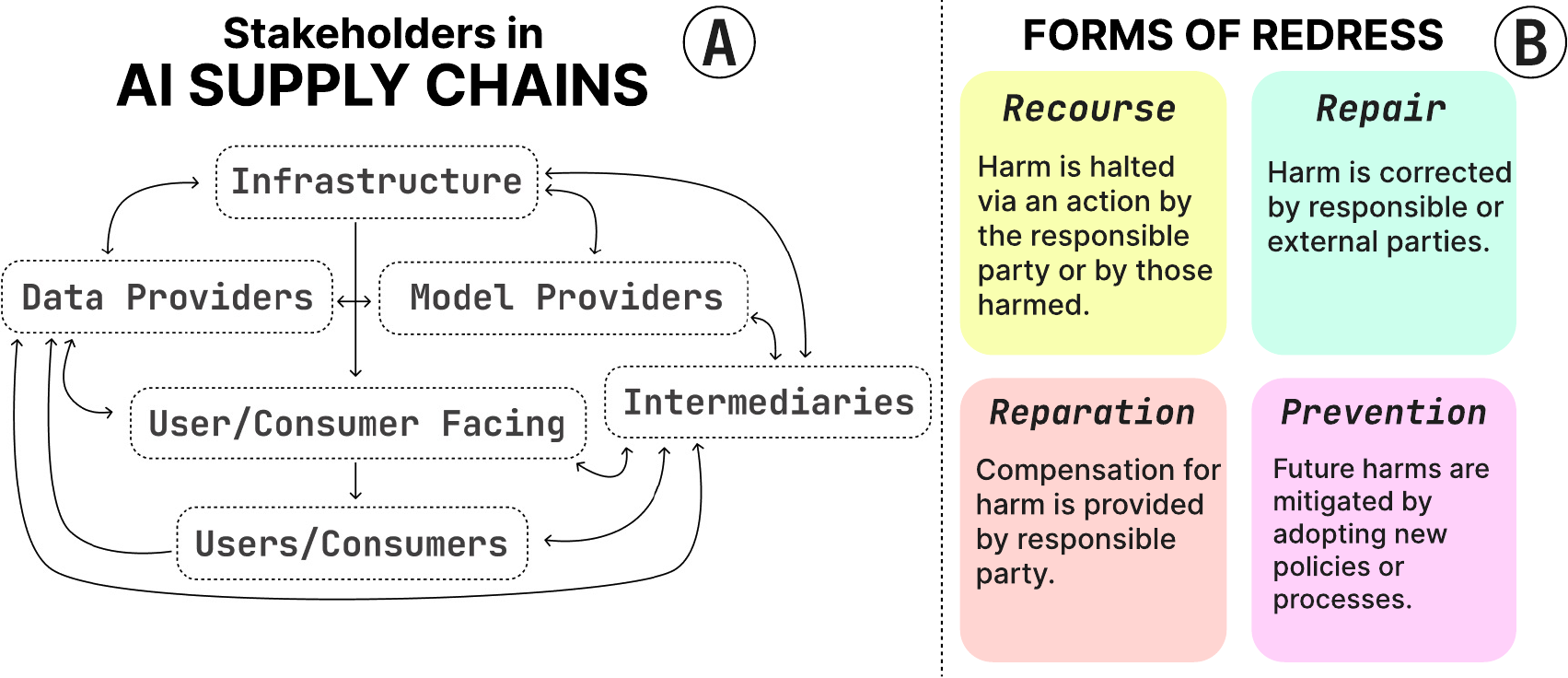}
    \vspace{-2mm}
    \caption{Stakeholders \& redress typology in AISCs. (A) An overview of stakeholder roles in an AI supply chain, including (1) Infrastructure, (2) Model Providers, (3) Data Providers, (4) Intermediaries, (5) User/Consumer-Facing, and (6) Users/Consumers. (B) A summary of the redress typology: recourse, repair, reparation, and prevention}
    \label{fig:responses}
    \Description{Overview: This figure comprises two panels, labeled (A) and (B). Panel (A) depicts the main stakeholder groups in an AI supply chain (Infrastructure, Model Providers, Data Providers, Intermediaries, User/Consumer-Facing, and Users/Consumers) and illustrates the flow of resources or influence among them. Panel (B) presents four color-coded boxes describing distinct forms of redress: Recourse, Repair, Reparation, and Prevention. Panel (A): Stakeholders in AI Supply Chains Visual Layout: A box for “Infrastructure” is positioned at the top; beneath it are “Data Providers” on the left and “Model Providers” in the middle. “Intermediaries” is to the right, while “User/Consumer-Facing” is slightly below and linked to both “Model Providers” and “Intermediaries.” At the bottom is “Users/Consumers,” which is connected back up to “User/Consumer-Facing” and “Data Providers.” Arrows: Arrows connect these stakeholder groups, highlighting how data, models, and services move or influence one another. For example, Data Providers and Model Providers exchange inputs and outputs, while User/Consumer-Facing services link to both the Intermediaries and the Users/Consumers. Purpose: This diagram shows the primary roles that organizations or individuals might occupy within an AI supply chain, emphasizing the interdependencies among them. Panel (B): Forms of Redress Visual Layout: Four colored boxes, each containing a short definition. Color \& Text: Recourse (Yellow box): Harm is halted via an action by the responsible party or by those harmed. Repair (Mint-green box): Harm is corrected by responsible or external parties. Reparation (Pink box): Compensation for harm is provided by the responsible party. Prevention (Lavender box): Future harms are mitigated by adopting new policies or processes. Purpose: Summarizes the four types of actions that can be taken to address harms within AI supply chains.}
  \end{figure*}
  
\section{Introduction}
Current efforts to understand what AI can and should do focus largely on desired end-user experiences and model outcomes. These efforts often miss an important factor shaping the use and consequences of AI: the AI systems we interact with are increasingly developed via fragmented ``AI supply chains'' through which multiple participants shape downstream use and effects. While AI supply chains (AISCs) enable firms to outsource otherwise prohibitively expensive technology, data, and expertise that support AI deployment, they differ in many ways from traditional supply chains. For one, by contributing data and layers of inference, the steps along an AI supply chain add more than just the functional and tangible components of downstream products. Downstream components may feed these additions \textit{back} into upstream processes in complex feedback loops, which lack transparency and control. Similarly, outputs from one AI system---a probabilistic determination---are increasingly used as \textit{inputs} to or \textit{training data} for another AI system. As these sampled distributions are based on likelihood, they lack a mapping back to real world examples or otherwise distinguishing characteristics specific to the model provider.
\footnote{While work on watermarking \cite{kirchenbauer2024watermarklargelanguagemodels} seeks to ameliorate this issue, it remains an open problem.} 
Thus traceability (or provenance) is challenged by both complex feedback loops and the ``non-injective'' property of distributions.
\footnote{A function (or system) is non-injective if it can map two or more different inputs (in the domain) to the same output (the codomain). In such cases, it's difficult to trace an outcome back to a unique source.}


Despite these challenges, the increased accessibility and efficacy of machine learning (ML) continues to encourage AI supply chains’ permeating ubiquity; such networks have grown common, complex, and diverse, expanding the number of contributors affecting downstream inference tasks \cite{griesch2023towards}, often with little documentation or disclosure of their upstream or downstream dependencies. While a simple AI supply chain may include only a base model (e.g., GPT-3 or ResNet-152) and domain-specific data (e.g., for properties of organic molecules \cite{xie2024fine}), AI supply chains now encompass multiple firms interfacing with the broader AI ecosystem and data capture, or directly with end-users, as in the case of Salesforce, Google, and Microsoft's recent integration of AI across platforms.  

While the AI supply chain phenomenon has become familiar, there is thus far no significant analysis that reflects the relative power and differential consequences resulting from participating and prospective stakeholders’ roles and interactions, nor has there been an accounting of how challenges presented by AISCs shape these interactions. 
Since the 1980s \cite{freeman2010strategic}, the concept of stakeholder has been widely adopted for just this purpose: to identify the actors with whom any organizational entity regularly interacts; i.e., any entity that has both an interest and influence in its activities. In this paper, we identify participating and prospective stakeholder roles, interests, and interactions as a first effort toward addressing the risks posed by AISCs.





We begin by recognizing two fundamental truths about AI supply chains.  First, the growth of AI supply chains is the result of a system that has provisioned massive computational capacity,
[ heretofore called ‘compute’],
 privatized funding and ownership, and substantive innovation and direction to a handful of actors. Second, 
the scale of AI adoption and model adaption (e.g., fine-tuning), along with the numerous AI services supporting AI use, means that AI supply chains vary substantially in organizational structure and (inter)dependencies. 
The sprawling, diverse, and rapid development of the AI industry and adjacent technologies has established a status quo that reflects little shared planning or centralized design through which affected stakeholders might work together toward common goals.  Instead, the industry’s growth appears \emph{organic} and \emph{unstructured}.

The appearance of organic emergence suggests that AI supply chains are not designed per se, and do not serve general---private or public---interests, for example, because they do not prioritize common incentives and shared language. 
They have emerged through complex webs of competitive and innovative transactions. 
As these networks have grown common, however, lack of design and human planning has led to assumptions of minimal responsibility, benefiting empowered stakeholders while introducing significant, even existential, risk to others.  

Our efforts are motivated by the following observation: AI supply chains are invented and implemented by humans with agency. They may, thus, be designed as a functional industrial ecosystem that accounts for the complexities, novelties, and risks of AI supply chains. 
Stakeholders are enabled or constrained by the dynamics within AI supply chains, and the current state of AI supply chains introduce information asymmetries and substantive disparities between stakeholders. By considering who participates in AISCs, what harms they face, where sources of harm lie, and how market dynamics inform the type and probability of remedies, our hope is to inform the design of functional AI supply chains. 

To this end, we characterize stakeholders by their role (participation) and interaction in AI supply chains. 
We identify mechanisms for harms that arise from the dynamics of these networks and the effects of AI, revisiting known mechanisms within the context of AISCs---false content, biased decision making, poor explanations and transparency, data scraping, security and privacy failures, and environmental effects---and compiling new mechanisms introduced to AI as a result of AISC adoption---reduced optionality, homogenization, dispersed responsibility. 
We then characterize stakeholder responses to realized risks with a typology differentiating among forms of redress: recourse, in which the harm is halted either through an action by the agent responsible or the harmed entity; repair, in which the harm is corrected; reparation, in which there is compensation for the harm; and prevention, in which the harm is mitigated before it occurs. 

Finally, we extend beyond simple enumeration to examine how AI supply chain variation---particularly in the accumulation and distribution of economic opportunity and power---shapes stakeholders’ capacity to respond to, and accept accountability for, these harms. Specifically, we apply our response typology and stakeholder characterization to a healthcare AISC example in three stylized markets---vertical integration, horizontal integration, and openly competitive (or ``free'')---to illustrate how the positioning and power of stakeholders within an AI supply chain influence the likelihood of both harm and response. 

\vspace{-2mm}
\subsection{Related Works}
Prior work has identified the systemic tendency for responsibility to be diffused in computerized societies, warning of a "many hands" problem where blame is easily shifted or lost altogether \cite{nissenbaum1996accountability,cobbe2020what_lies_beneath}. More recently, questions regarding how these same challenges are instantiated by AISCs have emerged \cite{cobbe2023accountability_supply_chains}. 
These works ascribe challenges to liability or accountability allocation,  scoping their focus to a specific interest (\cite{lee2023talkin} focus on copyright infringement and its legal implications, for example), or emphasizing the difficulty of allocation, though these efforts are often abstracted  from the context-specific challenges surfaced when a harm materializes (e.g., calls for transparency tooling \cite{widder2023dislocated,cooper2022accountability}'s).

In contrast, this work considers how stakeholder interactions and market dynamics challenge or otherwise support common actions neccessary in addressing technological and systemic failures: repairs, reparations, preventions, and of course, recourse. Specifically, we ask ``how do institutional or stakeholder arrangements enable or block redress?'' This necessarily requires us to consider that AISCs are not just emerging technical assemblages but structured economic arrangements---contractual, organizational, and market-based. Our efforts build upon \citet{pfeffer1978external}' theory of resource dependency, \citet{thompson2017organizations}'s work describing organizational interaction, and \citet{wrong1979power}'s analysis of power to consider how institutional logics such as risk outsourcing, liability avoidance, organizational boundary management, resource dependence, coordination frictions, and modes of integration inform the distribution of power and avenues of redress across AISCs. 

\section{Stakeholder Analysis}
\label{sec:stakeholder_analysis}
 Beginning from theories of the firm \cite{freeman1984strategic,freeman1990corporate, williamson1985economic}, stakeholder analysis has become a commonly used technique for identifying and managing challenges and risks arising from an organization’s activities and ongoing transactional relationships.  A stakeholder analysis of AI supply chains must consider not just those who build AI models but also those who “affect or [are] affected by the achievement in [an] organization’s objectives,”\cite{freeman1984strategic}. Rather than conceiving of the organization or firm as the focal center of an environment of surrounding stakeholders with simply dyadic or mutually independent relationships, stakeholder analysis recognizes that organizations produce consequences for their human and material environments, and therefore persist and thrive within highly interactive networks of mutual interdependence \cite{rowley1997moving, pfeffer1978external}, that may also include significant resource dependencies \cite{oliver1991strategic,frooman1999stakeholder}. Stakeholder analysis is both a form of social science— descriptive and predictive—and the basis of normative claims about how organizations ought to perform and how public policy ought to be made \cite{jones1999convergent, donaldson1995stakeholder}.

Simply naming categories of actors is not sufficient to differentiate stakeholders who can affect a firm or are affected by a firm, policy, or an AI supply chain. 
To be effective in identifying ``who or what counts'' in stakeholder analysis, however, one ought not “to weave a basket big enough to hold [all] the world’s misery” \cite{clarkson1994risk}.  
Instead, stakeholders should be identified as actors ``having something at risk'' in the relationship or system \cite{phillips1999stakeholder}. 

Prior work sorts stakeholder salience into three overlapping attributes: power, legitimacy, and urgency \cite{mitchell1997toward}. Power is the capacity to bring about intended outcomes in interaction with others \cite{weber1978economy, salancik1974bases, wrong1979power}. It is not a fixed trait but exercised through mobilizing resources—force, incentives, personality, moral leadership, expertise, or argumentation \cite{wrong1979power}—and carries no inherent moral valence. Legitimacy concerns the rightfulness of action, defined by social norms and often linked to legal standing. It is distinct from power; actors like terrorists or criminals may exert influence without legitimate claims. As Suchman puts it, legitimacy is “a generalized perception or assumption that the actions of an entity are desirable, proper, or appropriate within some socially constructed system of norms, values, beliefs and definitions” \cite{suchman1995managing}. Urgency captures the time-sensitivity and criticality of a stakeholder’s claim.

The primary boundary condition for identification as a stakeholder, of course, is whether an interaction actually affects the parties and what kinds of claims the actors can reasonably make upon each other. 
Here, we emphasize the term \textit{reasonably}; contestation and debate surround the central definitional and analytic questions concerning  stakeholder legitimacy, and what kinds of influence, interdependencies and dependencies characterize the stakeholders with interests in a particular organization or public policy, abound. 
As AISCs are new, largely unstudied and lacking in legal, social, or other forms of legitimization processes, we focus our efforts on urgency of claims, and power.



\section{Stakeholders}
\label{sec:stakeholder_roles}
Although relationships between an organization and stakeholders may be dyadic, segmented, and contractually specified in traditional supply chain analysis, many organizations within AI supply chains have relationships with multiple interdependent and often anonymous stakeholders, which produce nearly simultaneous influences and actions. To enable, structure, or govern such non-dyadic relations and information flows between entities in AI supply chains---for example, between model providers and downstream developers---requires a map of the stakeholders that are involved and their roles within the chain. Prior efforts to map the AI ecosystem include \citet{bommasani2024ecosystem} and \citet{mozilla2025publicAI}, though these either do not take a stakeholder framing or fail to account for AISC interdependencies. 

In following section, we describe ``who'' currently engages with AISCs---the stakeholders that participate in or are immediately affected by their use. Our categorization is the result of engaging with several datasets \cite{AIaaS_Supply_Chains_Repo, bommasani2023ecosystemgraphs}, and collecting organizational, service and product information from known AISC-participating tech companies. A more extensive characterization beyond what we provide here can be found in Appendix \ref{sec:extendedstakeholders}, and (inexhaustive) examples of vertical integration surfaced by these efforts can be found in Appendix \ref{sec:vertical}. In the following section, we describe the classification schema, illustrated in Figure \ref{fig:responses}, detailing the stakeholder categories by their contributions to the supply chain.

\subsection{Stakeholder Roles}
We identify the various contributions and functions of individuals, firms, and organizations within the AISC by their sector-agnostic role in the production and use of AI products and services. Actors may adopt multiple roles, such as contributing both underlying infrastructure, models, and products directly to users and consumers, or may play a single role, such as an end-user. In Section \ref{sec:casestudy}, we adopt an institutional lens to account for the relational complexities between firms in varied market settings. We focus this section on describing actors that are endogenous to AISCs (i.e., they have a direct stake in its function): those that are affected by, or contribute to, the chain. This is in contrast to exogenous actors such regulatory organizations, which may intervene in or study the AISC, but are still nascent in shaping its form or function. 

\vspace{2mm}
\noindent \emph{\textbf{Infrastructure Providers}} furnish fundamental technological tools and materials for building and operating AI systems. This includes chip manufacturing, data centers, and cloud services, all of which handle computation, storage, and networking at scale. Companies such as Amazon Web Services, Microsoft Azure, and Google Cloud offer broad suites of integrated products (e.g., compute instances, model deployment services, security protocols) that shape downstream development. Nearly all AI deployments depend on their products for compute, storage, and network capabilities, making them indispensable to downstream stakeholders. 

\vspace{2mm}
\noindent \emph{\textbf{Data Providers}} supply raw or processed information used to train, fine-tune, and evaluate AI models, as well as drive analytics. These providers can be platform owners (e.g., social media sites) that collect first-party data or aggregators specializing in assembling and labeling large-scale datasets. Data ownership and licensing are increasingly contested, especially when data arises from user-generated content or sensitive domains such as healthcare. Whether or not directly involved with the models they serve, data providers exert significant indirect influence over downstream outcomes. 

\vspace{2mm}
\noindent \emph{\textbf{Model Providers}} develop AI models, offering them to downstream stakeholders as standalone products or APIs. Providers may release complete models for public use (e.g., EleutherAI, Allen Institute for AI), allow restricted API access (e.g., OpenAI, Meta), or form exclusive partnerships with other firms. The scale of models, the resources they require, and their licensing terms shape how easily downstream developers can adopt or adapt them. Currently, a handful of well-resourced organizations dominate the creation of cutting-edge models. Smaller actors focus on niche applications and build upon open-source projects. In many cases, \emph{platform effects} arise: model providers that bundle developer tools, deployment pipelines, or specialized features can create lock-in for downstream developers. Applications built around a particular model’s idiosyncrasies may be costly to ``transplant'' to other ecosystems, thereby reinforcing concentration and incentivizing continuous adoption of a single provider’s platform.

\vspace{2mm}
\noindent \emph{\textbf{Intermediaries \& AI Services}} offer specialized functions---such as data processing, model fine-tuning, monitoring, or deployment---that fit between existing data, models, and infrastructure providers. For example, a firm offering automated data labeling tools integrates upstream raw data with downstream model training pipelines. They typically do not interface directly with end-users; rather, they plug into the pipeline to refine workflows, automate MLOps tasks, or assemble specialized services. As dependencies layer on top of one another, these intermediaries can significantly increase the complexity of AI supply chains. However, in many cases, infrastructure giants supply these ancillary services.

\vspace{2mm}
\noindent \emph{\textbf{User-Facing}} entities deliver AI’s functionality directly to end-users through interfaces, applications, or APIs. Examples include chat platforms (e.g., Character.AI) that wrap large language models in accessible tools, or all-in-one services (e.g., Microsoft Office Co-Pilot) that embed AI across user workflows. They often incorporate additional layers—such as prompt-engineering guidelines or safe-mode settings—to adapt a core model for particular uses. User-facing providers depend on model providers and infrastructure services behind the scenes, but they sit at the boundary between AI systems and the public, and they heavily influence how models are presented, interpreted, and used in practice. 

\vspace{2mm}
\noindent \emph{\textbf{Users \& Consumers}} (individuals and organizations) engage with AI systems---consciously or passively---through search engines, recommendation platforms, enterprise tools, or specialized applications. Whether they pay for or freely access AI services, they often have limited visibility into upstream processes. However, large enterprise users 
can exert notable influence over model and infrastructure providers, and individual consumers commonly shape AISCs through their aggregated behavior.  As in any market or major social field, collective demand influences the strategies of upstream suppliers, who may adjust offerings individual or in partnership with other to serve current or anticipated user interests, or regulatory environments, should they develop. Notably, users’ expertise, use volume, and economic situations vary widely, making consumers a highly heterogeneous but central influence in AISCs.

\section{Mechanisms of Harm in AI Supply Chains}
\label{harms}
AI poses sizeable and significant risks, as prior work has highlighted, including reifying existing power imbalances while exacerbating social inequalities \cite{suresh2024participation}. Descriptions of AI risks have typically focused on the risks to individuals and specific demographic groups. For example, \textit{allocative} risks---in which algorithmic bias may lead to resource or opportunity distribution (e.g., in hiring or loan approvals), potentially disadvantaging specific 
social groups---is distinguished from \textit{representational} risks---wherein algorithms may reinforce stereotypes or discrimination against specific groups, and which is turn contrasted with \textit{explanatory} risks, in which individuals are unable to understand how decisions were made \cite{metcalf2023taking}. 

Recent reports from government agencies have expanded upon these definitions to frame investigative efforts and shape priorities for possible AI regulatory initiatives. For example, the U.S. National Institute for Standards and Technology (NIST) Artificial Intelligence Risk Management Framework \cite{ai2023artificial} offers a set of risks that range from increased ease in developing weaponizable products (chemical or otherwise), to data privacy and AI's environmental impacts. These characterizations do not account for the risks faced by organizations, and, perhaps more critically, do not (yet) encompass specific risks surfaced by AI supply chains (though \citet{ai2023artificial} points out that risks \textit{do} exist). 
To this end, we recapitulate mechanisms of AI-induced or exacerbated harms in Table \ref{table:harms}, noting possible consequences for individuals, organizations, and systems. Distinguishing between such scales [ individuals, organizations, systems ] emphasizes that how harm is experienced, attributed, and addressed will vary across the setting in which the harm occurs, and mirrors common social, organizational, and regulatory science groupings \cite{huising2018nudge,stinchcombe2013social}. \emph{Our aim is to articulate how known harms are expressed through or compounded by AISC structures.}

\begin{table*}[htbp]
\centering
\renewcommand{\arraystretch}{1.1} 
\setlength{\tabcolsep}{3pt}       
\caption{Mechanisms of Harm in AI \& AISC}
\label{table:harms}
\begin{tabular}{p{4cm}p{4.25cm}p{4.25cm}p{4.25cm}}
\toprule
\textbf{\small Mechanism of Harm} & \textbf{\small Individuals \& Groups} & \textbf{\small Organizations} & \textbf{\small Systems} \\
\midrule

\textit{False Content} &
\small Harassment, misrepresentation, misinformation &
\small Harm to reputation and autonomy &
\small Reduced social trust \\
\midrule

\textit{Biased Decision Making} &
\small Economic harms, discrimination, representational harms &
\small Undesirable performance, economic harms &
\small Magnification of systemic weaknesses \\
\midrule

\textit{Poor Explanations} &
\small Lack of liability or recourse, explanatory harms &
\small Minimal liability pathways &
\small Reduced traceability \\
\midrule

\textit{Data Scraping} &
\small Privacy violations, IP infringement &
\small Economic harm &
\small Reduced traceability, unclear ownership \\
\midrule

\textit{Security \& Data Privacy} &
\small Privacy violations &
\small Data breaches &
\small Risk of system failure \\
\midrule

\textit{Environmental} &
\small Physiological, economic harms &
\small Economic &
\small Environmental degradation \\
\midrule

\textit{\textbf{Diffused Responsibility}} &
\small Minimal liability pathways &
\small Minimal liability pathways &
\small Reduced trust, lack of standards \\
\midrule

\textit{\textbf{Reduced Optionality}} &
\small Economic harms, loss of choice &
\small Economic harms, harm to autonomy &
\small Reduced innovation \\
\midrule

\textit{\textbf{Homogenization}} &
\small Loss of choice &
\small Security vulnerabilities &
\small Reduced resilience \\
\bottomrule
\end{tabular}
\end{table*}

While threats to national security and public safety are critical, our interests in this work lie with the day-to-day concerns of stakeholders. Our list compiles previously collected mechanisms of harm that may be explicitly affected by the introduction of AISCs. Mechanisms that are afforded by AISCs, and thus are in many ways new to AI, are in bold. In Table 1, we summarize how AI supply chains can add additional complexity to known AI risks and may introduce new mechanisms of harm.
\footnote{Not all AI-related harms are explicitly reshaped by AI supply chains. \textit{Human-machine emotional entanglement}, in which individuals form deep attachments to AI systems, \textit{overreliance}, where users or organizations become excessively dependent on AI to the detriment of their own competencies or critical thinking, and the \textit{normalization of AI-induced harms}, in which stakeholders gradually accept detrimental AI outcomes as unavoidable, are important concerns for AI broadly. However, these particular mechanisms emerge from broader human and organizational relationships with AI and persist even in simple AI deployments, thus are excluded.} 
Engineered systems like AISCs are not inherently static---there will be updates, changes, and new priorities adopted across any given AISC. Performance variability is a natural result of complex systems \cite{hollnagel2006resilience}, and this, along with the priors already posed in AI development, presents risks for organizations, systems, and individuals. 
As stakeholders interface with AI supply chains, these risks will be realized as harms. 

\subsection{False Content} False content, for example, has led to misrepresentation, misinformation, and harassment---with consequences to employment, reputation, and bodily autonomy \cite{umbach2024non,saenz2024let}. Customers of banks that adopt voice identification verification, for instance, face a growing security threat from voice cloning \cite{morrison2023aivoice}. Organizations experience reputational and economic consequences from misinformation both indirectly and directly:
\footnote{ AirCanada's recent arguments regarding chatbot liability following customers being misinformed on company policies, and the effects of fake information about explosions at the  White House causing the S\&P 500 to lose more than \$130 billion in market capitalisation \cite{karppi2016social} offer two examples.}:
 misinformation shapes ``misperceptions'', which in turn affect firm valuations \cite{arcuri2023does}. Systemically, widely circulating AI-generated false content has normalized public skepticism concerning not only individual factual claims but also the presentation of truth per se, feeding widespread loss of trust in public institutions, political processes, and specific organizations  \cite{kreps2022all}. Consider ‘swat’ attacks as another example of misinformation. In a swat attack,  a person intentionally creates a synthetic, realistic voice, and calls the police to falsely report a bomb in a school, hospital or other public space. The police respond by closing down schools and hospitals with accompanying community disruption. Some such attacks also target individual homes or organizations, generating and escalating fears, rather than confidence, in the protective agencies of the police. \cite{hutiri2024not}.


AI supply chains can amplify the avenues of harm caused by false content in several ways: first, other AI systems or AI agents may prompt generation of content or deepfakes, escalating the scale of production with minimal friction or human overview. In this case, an agent may interface with a generative model to produce the false content. Second, false generated content may be adopted into a training set. For example, content may be incorporated into platforms or search results. This may emphasize its rhetorical force and legitimacy to individuals, particularly when it is challenging to critique or evaluate. The ease of generation and the interdependencies of AI systems may then enable said content to proliferate across platforms, systems, and modalities, reinforcing the false material. As the misinformation mushrooms, it may be incorporated directly into training sets through leaky avenues of data collection. Similarly, the content may be incorporated into \textit{prompts} that are used to reinforce a model. In banking, a voice clone might be used to interact more frequently with a bank than the true customer. With time, these samples might reinforce the clone as ``real'' and the person as ``fake''. If not salient to model developers or their upstream and downstream partners,  content may thus be re-generated, amplified, or may subtly reinforce undesirable behavior. 

\subsection{Biased Decision Making} Biased decision making has economic and material consequences ranging from misdiagnoses to inequitable bail decisions. Bias in one AI component can propagate across the AISC: this is because \textit{what is learned upstream will have effects downstream}. In other words, upstream decisions shape downstream outcomes \cite{hopkins2025ai}. AI supply chains have already been shown to bias housing recommendations \cite{liu2024racial}, and have ample opportunity to produce similar inequalities in other material and social settings. And the influence of upstream decisions may be exterted in less direct ways. Consider a retail company that contracts a chatbot vendor for customer support. The upstream model's safety mechanisms lead to refusals for inquiries that are reasonable within the retail setting---cleaners, BB guns, or sex toys---but that are topically not acceptable to the model provider, exacerbating customer frustration and churn. Alternatively, upstream and downstream datasets may interact in unexpected ways. For example, overlap in datasets introduced at different points of an AISC may lead to calibration errors in downstream systems that disproportionately affect certain subpopulations. When deployed in regulated industries like finance or healthcare, the resulting adverse action may trigger regulatory intervention. Without targeted disclosures of system design decisions, downstram developers will struggle to recognize biasing effects across the AISC.

\subsection{Poor Explanations} Poor explanations challenge the fundamental transparency necessary for supply chain and market transactions. Many AI systems, especially those based on deep learning, function as ``black boxes,'' where even model developers struggle to interpret decision processes \cite{lipton2016mythos}. Proposed solutions abound, including local explanations, saliency mapping, and mechanistic interpretability, though poor robustness or high costs of computing can render them unreliable or intractable at scale.  And challenges in understanding the resulting explanations can be significant, even in inherently explainable machine learning architectures \cite{hong2020human}. Such opacity hinders users' understanding of why a decision was made, in turn challenging responsibility attribution to either the developers, deployers, or the AI system itself \cite{wachter2017right}. This has immediate consequences for how harms can be responded to, and is a theme we revisit in our case studies.  When AI systems produce poor explanations, it is challenging to identify and address erroneous aspects of algorithms or training data. 

This raises legal and ethical concerns when individuals are adversely affected by AI decisions
 without sufficient explanation to motivate redress.
 Affected individuals cannot effectively challenge or appeal AI-driven decisions without an adequate understanding of the underlying reasoning. Nor can they learn from or target their behavior to change an outcome. An AI-driven hiring tool may reject candidates without providing meaningful feedback, leaving applicants unable to address or understand deficiencies, or an automated loan denial might not present the actual feature that led to rejection. In both settings, existing regulation tends towards requiring an explanation.

AI supply chains further exacerbate issues of traceability and explanations, in part because there are few to no standards on how to design for these features across organizational or compositional boundaries. If an output is the result of multiple AI systems, should an explanation be a composition of individual explanations? Is the last-most model's explanation sufficient? Would any of these forms of explanation be accurate? Regardless of efficacy, AISCs are frequently the result of outsourcing to other organizations. Dependencies on proprietary AI services or upstream AI models can prevent traceability beyond a single entity \cite{hopkins2025ai}. The large scale and domain-agnostic manner in which AI models operate heightens the stakes: any errors or harms are ``liable to reoccur across use cases'' \cite{suresh2024participation}. We know from historical experience since WWII, in the creation of the global monetary system and financial markets, that a rules-based regime 
(even in a vast system of high-speed highly technical transactions)
 depends on differentiated levels of secure transparency. Yet standard mechanisms in which to audit or trace \textit{across} AISCs do not exist.

\subsection{Data Scraping} 
Data scraping has fed the scale of growth for AI models, companies, and AI supply chains. While  scraping user data from websites can infringe on privacy rights and regulations like GDPR, CCPA, or HIPAA, major legal tensions have centered on whether AI developers bear liability for training their models on copyrighted datasets, even if those datasets are publicly accessible online. Merely scraping publicly accessible data does not exempt developers from copyright obligations; instead, courts have begun scrutinizing if the use of such data is transformative and proportionate under fair use doctrine. If a model's outputs closely replicate original works, it may constitute infringement. Under doctrines like contributory and vicarious liability, companies that facilitate scraping could be held accountable for downstream violations \cite{lee2023talkin}.

The scale of data scraping has led to frictions in vetting the legality of data sources and obtaining necessary licenses for proprietary content. In response, a number of high profile, exclusive licensing by model providers with news and other literary sources, including between Meta and Reuters \cite{reuters_meta_ai_2024}, OpenAI with Wall Street Journal and AP News,  Google Gemini with AP News \cite{zaidi2025associated} (at least the former with the intention of integrating real time, vetted news sources for rapid model updates), and Microsoft with HarperCollins \cite{farooque_microsoft_harpercollins_2024}. Even outside explicitly protected content, licensing with social media companies (when not already one) for training data has also become relevant, as shown by Google and Reddit's agreement in 2024 \cite{hamilton_reddit_google_2024}. These contracts are increasingly bidirectional in services rendered: for example, Axios now provides stories and training data to OpenAI, which in turn serves its technology for creating and distributing journalism---and funding local newsrooms \cite{axiosgoogle}. It is unclear how these dynamics affect the landscape of journalism, but the consolidation of licensing agreements does not necessarily preclude webscraping and raises questions of power concentration and homogenization of literary and news content.

\subsection{Security \& Data Privacy}
At the organizational level, security and data privacy are concerns that are only escalated by AISCs. Consider the costs of data breaches and cyberattacks against healthcare, finance, retail, manufacturing, energy and government agencies have been estimated to be over \$10.5 trillion \cite{fine2023cyberrisk, henning2014cyberattacks}. While large firms are investing in new forms of cyber security, small firms struggle to mobilize affordable security strategies.  In 2023 alone, Samsung, 23 \& Me, Walmart, DuoLingo, Microsoft, T-Mobile, Discord, Eye4Fraud, Chick-fil-A, Verizon, and Google Fi among many others 
reported data breaches exposing the financial and personal information of users, clients, and employees. Access to this information creates substantial risks, including credential-stuffing attacks, phishing schemes, and unauthorized account access. While the risks of cyber-attacks are known, AI supply chains require us to revisit how they might surface. Data poisoning and security breaches via open models grow more common \cite{chaudhuri2024securing}. Linkage attacks \cite{andreou2017identity, merener2012theoretical}, where multiple sources of information are used in corroboration to identify individuals or groups, may be re-appropriated to AI supply chains. As new AI models and datasets are introduced into the AI supply chain, the possibility of such attacks may grow, with such consequences as de-anonymization, identity theft, increased robocalls
\footnote{Case in point, the FCC recently fined Lingo Telecom \$1 million for transmitting AI-generated robocalls that imitated President Joe Biden's voice \cite{fcc2024doc404951}.}
\cite{fcc_robocalls_fine} and fraud \cite{ic3_elder_fraud_2023}, often targeting older, vulnerable populations .
\footnote{Nearly half the complainants to the FBI's internet crime complaint center reported to be over 60 and experienced 58\% of the losses (nearly \$770 million). For context, only \~17.3\% of the US population is over 65 \cite{uscensus2023population}.}




\subsection{Environmental Impacts}
The environmental impacts of increased AI adoption, exacerbated by AISCs, are a growing concern, highlighting a current rift between AI-driven economic growth and \textit{sustainability}. On January 15, 2025, Google announced that Google Workspace was integrating its advanced AI capabilities directly into Business and Enterprise plans. The announcement described  how ``AI enhancements'' for business services, including Gmail, Docs, Sheets, Meet, Gemini and Chat, are adopted and priced \cite{google2025workspaceai}; rather than opt-in, customers could not opt out. Each customer on a workspace plan pays \$2 a month more to benefit (a relative bargain for an otherwise more costly product like Gemini). This expansive shift reflects common decisions made by tech-forward companies to integrate AI across and between products, and is tractable at this scale because of existing vertical integrations.  While there are ways to reduce how resource-heavy a given model is (e.g., small models are less computationally intensive), implementing AI where AI tooling \textit{may not be wanted} does not align with collective stakeholder interest. These systems are developed and deployed using resources that require meaningful water, electric, and mineral consumption; yet AISCs encourage AI use in broader contexts than previously seen. As norms in AI-use shift rapidly, reports of new data center developments---increasingly owned by cloud, hardware, and model providers---and their burden on local communities abound \cite{valdivia2024supply}.




%




\subsection{Diffused Responsibility}
Diffused responsibility arises when development and use of an AI system are spread across multiple teams and organizations, and no single entity holds full accountability for its outcomes. This is particularly pronounced when core AI components are externally sourced or licensed; an update to a base model or a training dataset might propagate new biases or errors down to a product that integrates it, without the deploying organization’s knowledge or ability to reverse the change. As control is partitioned among various entities, each upstream or downstream step in the chain can obscure who is liable for harmful outcomes. A downstream organization might, for instance, claim that any problematic changes stem from the upstream provider, while the upstream provider could insist that those issues are rooted in how the model was integrated into the downstream product. With few standards of practice or robust record-keeping across the chain, tracing the precise point of failure---or the party with the power to remediate it---becomes difficult. Together, these dynamics create a ``nobody’s fault'' environment wherein accountability is diffuse, and each participant plausibly redirecta blame to other parts of the supply chain.

This fragmentation leads to practical consequences for resolving harms. Individuals who experience harm may be passed from one technical support channel to another in a fruitless attempt to find redress. Meanwhile, organizations themselves risk compounding liability; not only may they bear the fallout from unmet contractual obligations or compliance violations, but they also shoulder reputational damage from problems that originate elsewhere in the supply chain. Systemically, the lack of clear accountability erodes trust in AI-driven processes and complicates oversight. 
AI supply chains can thus obscure lines of accountability, introducing significant friction in rectifying harms.

\subsection{Reduced Optionality}
Reduced optionality describes the gradual erosion of choice in AISCs—both in selecting AI services and in negotiating how they are provided. This reduction in optionality stems largely from the consolidation of AI capabilities, especially large foundation models, within a small set of upstream providers. It's a common phenomenon in traditional supply chains but is new to AI. Only a few companies possess the resources (massive datasets, specialized hardware, highly skilled researchers) to develop and maintain billion-parameter-scale models.
As more organizations rely on these incumbents’ APIs or toolchains, the ``lock-in'' effect is compounded.


High switching cost associated with AI integrations can drive reduced optionality. Unlike generic software, large-scale model use can be tightly coupled with unique feature dependencies (e.g., upstream embeddings), which along with architecture and tuning decisions, introduce priors downstream developers may not have access to.
Migrating to a new provider may require (costly) retraining and evaluation or adapting core components of an application, rebuilding engineering workflows, and retraining staff. These logistical burdens may deter even well-resourced organizations from switching.
Adding to the difficulty, many model providers offer ``bundled'' services---analytics, developer tools, proprietary optimization frameworks---that become integral to a company’s operations. Exiting the ecosystem may mean overhauling multiple interdependent layers of technology. 

From a legal perspective, reduced optionality may introduce frictions when negotiating contracts that clarify liability or guarantee acceptable performance. Model providers can use opaque or non-negotiable terms that disclaim responsibility, leaving downstream entities to absorb both reputational and regulatory risks. Lack of meaningful competition means there is scant market pressure to offer more transparent or balanced liability clauses. Should a model malfunction due to bias, security breaches, or even simple outages, downstream adopters may have no viable path forward; industries as diverse as e-commerce, healthcare, finance, and education may find themselves simultaneously reliant on the same small set of providers with few redundancies, magnifying any single failure into a potential system-wide crisis.



\subsection{Homogenization} 
Homogenization refers to the convergence of AI outputs, processes, or architectures toward a narrower range of possibilities, to the detriment of innovative competition and resilience. In the context of AI supply chains, homogenization can manifest at multiple levels. First, companies may repeatedly adopt the same tokenizers, base models, training data, or architectures, leading to a limited set of downstream applications that are structurally and behaviorally similar. Prior work on algorithmic monocultures has illustrated the negative consequences of this phenomenon \cite{kleinberg2021algorithmic, bommasani2022picking}, which occurs in part because organizations favor ``proven'' AI systems, or because a small handful of large providers dominate the market. At scale, such convergence can reduce competition, restrain creative exploration, and shape user experiences toward a uniform and potentially stale set of interactions \cite{jo2025homogeneousalgorithmsreducecompetition}. In an AISC, repeatedly outsourcing or licensing these same base models further concentrates design choices, propegating this similarity to downstream products. 

In AI components, a particularly acute form of homogenization emerges when AI-generated content is re-ingested as either training data or as inputs to another AI system. One example of this is mode collapse, well-studied in GANs \cite{goodfellow2020generative} and increasingly in LLMs \cite{shumailov2024ai}, wherein a system’s outputs converge to a reduced set of high-probability variants. This can result in the long tail of more creative, infrequent, or context-specific expressions effectively disappearing over time. Such homogenization shrinks the expressive capacity of AI and increases systemic fragility: if a single model or dataset harbors inherent strong biases or vulnerabilities, these flaws may be disseminated throughout an entire ecosystem. Thus homogenization, especially in complex AI supply chains, may diminish both resilience and representational breadth, reinforcing shared error modes and limiting the diversity of AI-driven innovation. 

\section{Redress Typology}
\label{redress}

Risk that surfaces as a result of adopting new technology is in many ways ``old hat''; this is the focus of risk and resilience literature in both engineering and management, which seeks to reduce the dislocations and inefficiencies that result from systemic, organizational, or technological transitions \cite{hollnagel2006resilience, morison2016men}. These dislocations are risks, or potential harms, to participants of a system. Unfortunately, harm---whether economic, emotional, physical, or reputational---is not always preventable. Instead, it is often in the subsequent \textit{response to} said harm where intervention is possible. 

Defining, interrogating, or implementing redress is the focus of multiple disciplines that operationalize various schemas to identify what is appropriate or valuable in distinct societies \cite{macintyre1984after}. Terms such as recourse, reparations, restitution, resolution, recovery, repair, and redress (among others) often hold similar meaning in popular discourse, yet distinct referents to specialized and professional audiences \cite{rawls2020theory, gold2014theory, perrow1984normal, abbott1988system}. The substance of tort law is concentrated on just these topics, identifying the grounds of responsibility and response for injuries. Insolvency law similarly establishes procedures such as credit recovery claims, where the distribution of assets is dependent on judicial determination of stakeholders’ status and injury in relation to the insolvent party. 
Restitution, and reparation, typically encompass attempts to address systemic harms by restoring and/or compensating for losses---e.g., as reparative act to historical oppression (though reparations has broader applications in legal justice) \cite{de2008handbook,rothstein2017color}. In the scope of AI, recourse is defined as the actions a prediction recipient can take to reverse a given decision \cite{upadhyay2021towards}, but is referenced beyond the domain as steps taken to \emph{fix} an erroneous outcome by involved party/ies \cite{ghent2011recourse, solomon2011civil}. Simply recognizing responsibility has, by some, been ascribed as a form of redress wherein a perpetrating party acknowledges responsibility \cite{Tirrell2013, gibney2001status, govier2002promise}. We posit that recognition of a harm---if not necessarily by the responsible party---is a prerequisite to any act of redress, and do not include it in our typology.

Despite the variation in terminology and application, there are limited forms in which to seek redressment. While work in law and machine learning has described actions that can be taken when a harm occurs, there is no schema collating the diversity of perspectives and positions. 
To this end, we contribute a typology distinguishing avenues of redress 
(shown in Figure \ref{fig:responses}):
\textbf{recourse}---in which the harm is halted either through an action by the agent responsible or the entity that is harmed; \textbf{repair}---in which the harm is corrected with the intention of returning the harmed party to a prior state; \textbf{reparation}, in which there is compensation for the harm; and \textbf{prevention}---in which the harm is proactively mitigated before it occurs. 
Our typology is a logical synthesis produced through conceptual comparison and alignment---for example, linking ``reparation'' in transitional justice to ``compensation'' in consumer harms. 

Differences in status are most resonant when we consider how a harm experienced by one actor is responded to by a responsible party \cite{mackinnon1989}. 
Stakeholders can and will have unequal bargaining positions. Downstream entities may have little leverage when confronted by an integrated or specialized model provider that holds a monopoly or a vital data resource. In contrast, if more competitive settings or regulatory oversight exists, injured parties will have more influence to demand timely repairs or seek reparation.
Whether a redress is achieved in an AI supply chain is determined by two considerations: \emph{consensus}, whether necessary actors agree upon (or are effectively compelled to accept) a proposed remedy, and \emph{achievability}, in that the redress is [physically, technically, legally] possible given the involved parties' and systems' limitations.
Even if a technically viable fix exists, entrenched power and perceived urgency with regards to stakeholder relations influences how quickly and fully any redress will be pursued. If halting a harm (recourse) or compensating for damages (reparations) is theoretically warranted, a stakeholder may still lack the technical or financial standing to implement the fix, or a suitable legal framework to enforce it. 

As discussed in Section \ref{sec:stakeholder_analysis}, stakeholders may play multiple roles in AISCs. Power differentials, the extent and mechanism by which harms are realized or perpetrated against stakeholders, and the urgency of demands thus reflect these variations---and inform the type, extent, and feasibility of redress. In the following section, we consider how redress is informed by the positionality between redress seeker and enactor (i.e., consensus and achievability), characterizing how forms of integration might enable or otherwise disable redress between stakeholders in an AISC.
\section{Market Configurations}
\label{sec:casestudy}

Concentration in AI supply chains creates inefficiencies, instabilities, and significant power asymmetries that exacerbate the existing risks posed by AI. Power differentials shape stakeholder relationships in ways that \textit{inform the forms of redress} provided in response to harms. Redress will vary across AI supply chain contexts. A response may not be technically or fiscally achievable, or be constrained by poor traceability, or the stakeholders involved in the harm may not reach consensus on what is a reasonable form of response. In other settings, the power differentials among involved parties, and the significance and urgency of a harm may affect whether and how consensus is reached. In this section, we consider how a set of harms across a single AISC is responded to when different market configurations are at play. 
We do so to understand the distinct challenges that concentrated power (i.e. oligopoly, monopoly, or monopsony) poses for AI supply chains, and to inform alternatives in the future. 
We focus on three market configurations: vertical integration, horizontal integration, and the free (or openly competitive) market.

\begin{figure*}
    \centering
    \includegraphics[width=.98\linewidth]{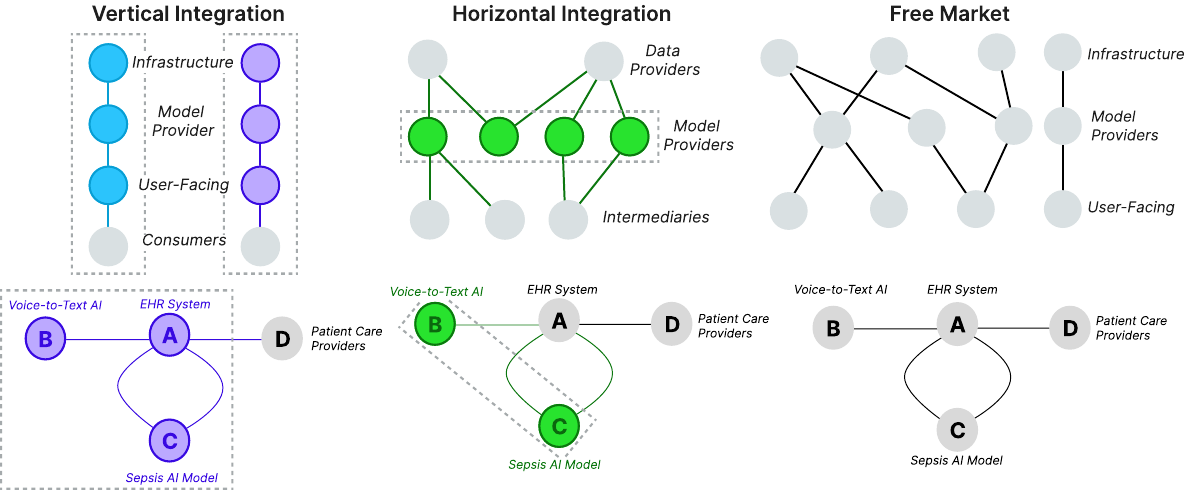}
    \caption{General AISC and sepsis model AISC are shown under vertical integration, horizontal integration, and the free market.}
    \label{fig:marketsss}
    \Description{Top Row. Vertical Integration (Left): A stacked column of colored circles representing distinct stakeholder roles (Infrastructure, Model Provider, User-Facing, and Consumers) enclosed in a dotted box. Each circle is vertically aligned, indicating that a single firm spans multiple levels of the supply chain. Horizontal Integration (Center): Several circles in two groups, each group connected via lines that indicate shared or aggregated control at one level (e.g., multiple Data Providers on one layer, multiple Model Providers on another). Different shades of green denote the horizontally integrated entities. Free Market (Right): A network of gray circles connected by black lines, with no single entity dominating. Each circle stands for a different firm or stakeholder role (Infrastructure, Model Providers, or User-Facing), illustrating a more diffuse or competitive market structure. Bottom Row. Vertical Integration (Left Sub-diagram): Four labeled nodes: B (Voice-to-Text AI), A (EHR System), C (Sepsis AI Model), and D (Patient Care Providers). Nodes A, B, C are shown in a single integrated color scheme (purple and blue), with loops indicating interactions between them. Node D sits outside this integrated block. Horizontal Integration (Center Sub-diagram): Four labeled nodes in green/gray: B (Voice-to-Text AI), A (EHR System), C (Sepsis AI Model), and D (Patient Care Providers). B and C are highlighted in green, showing that they belong to a horizontally integrated firm at the “Model Provider” layer, while A (EHR System) and D (Patient Care Providers) remain external. Free Market (Right Sub-diagram): Four gray nodes: B (Voice-to-Text AI), A (EHR System), C (Sepsis AI Model), and D (Patient Care Providers). Each node is separate, connected only by simple lines to indicate a flexible, market-driven arrangement with no dominant integrator.}

\end{figure*}

\vspace{2mm}
 \noindent \textbf{\textit{Vertical Integration.}} In a vertically integrated market, a single firm controls the entire production chain---from extracting raw materials to manufacturing, distribution, and final delivery---thereby creating a monopoly. In contrast, our setting involves a vertically integrated firm operating within a larger market (Figure \ref{fig:marketsss} A). Though this firm manages its own end-to-end supply chain, it competes alongside other firms (which may also be vertically integrated). In such a market, one firm does not necessarily dominate all activity; instead, there may be a small number of powerful, vertically integrated competitors. 
 A classic example is the American automobile industry in the mid-20th century, when a handful of dominant, vertically integrated automakers each managed its own supply chain while still competing with one another.

\vspace{2mm}
\noindent \textbf{\textit{Horizontal Integration.}} A horizontally integrated firm (Figure \ref{fig:marketsss} B) single-handedly (or by merging with other groups) expands its market power and offerings to control particular form of provisions in the supply chain level. In AISCs, this may be any stakeholder role (such as model providers), and it may also be a stakeholder role in a specific domain (such as computer vision model providers). Though the rest of the supply chain remains competitive, the firm maintains control over the production process it operates. 
This contrasts with a vertically integrated firm that spans (and does not necessarily control) the entire supply chain. Prominent examples include Carnegie Steel (now U.S. Steel) and Rockefeller's Oil Company (ExxonMobile, among others).

\vspace{2mm}
\noindent \textbf{\textit{Free Market.}} In a free market (Figure \ref{fig:marketsss} C), many small firms produce substitutable goods, and no individual firm is able to control pricing. Prices are determined by market supply and demand, and firms act as price takers. While barriers to entry or exit might influence firm costs, in principle costs derive from the particular capital, organizational, and material circumstances of particular firms who operate with comparable knowledge of market conditions. 
The variation present in the restaurant industry is a prominent example of a highly competitive free market.

\section{Healthcare Case Study}
Consider a hypothetical AISC for healthcare diagnostics, illustrated in Figure \ref{fig:marketsss}.
There are four stakeholders: (A) an electronic health record (EHR) platform, acting as both data provider and intermediary between other stakeholders, (B) a model provider that serves a voice-to-text AI model for transcribing patient visits and doctor dictation, (C) a model provider of an AI sepsis product used for supporting patient triage, (D) hospitals, and expert and non-expert users---that is, doctors and their patients. In this setting, the sepsis model from C is under a trial run with several hospitals (D) under short-term contracts where the hospitals renew based on continued performance. The model interfaces with the EHR system from (A), which provides aspects of patient health to inform their risk rating. The hospital is also contracted with the voice-to-text provider (B), which is conveniently integrated into the EHR system (for now, we will put aside the fact that (B) may also have upstream dependencies, e.g., may be fine-tuned from an upstream model). When a doctor visits with a patient, the patient's records are updated by (B) to include changes from the visit.

The EHR system provides patient information to the sepsis model and is updated by the voice-to-text model. When (B) updates its voice-to-text model, improving overall performance, the update introduces an uncaught bias: now, the model is worse at correctly translating a small number of medical terms for doctors with certain accents. This distribution shift, though small, has cascading effects. The sepsis model (C) no longer accurately reflects patient sepsis risk in some hospitals---leading to reduced responsiveness by doctors to high risk cases and, critically to (C)'s interests, causing several hospitals to not renew their contract, reducing profits.
The voice-to-text provider has fiscally impacted the sepsis model provider by changing a product, with cascading consequences for patient health and hospital reputation.

There are several mechanisms of harm that inform this scenario: biased decision making by an upstream provider's model has shifted aspects of data that the downstream sepsis model provider uses as input, while dispersed responsibility and poor traceability may make it difficult for either A, B, C, or D to isolate the cause of the sepsis model's failure. Without proper incentives across stakeholders, it may not be possible to uncover what initiated the failure. If steps to isolate and respond appropriately are not taken in a timely manner, the effects may propagate across the AI supply chain: biased outputs from the voice-to-text model may eventually be incorporated into future training data, or the sepsis model's poor performance may be reinforced through iterative updates. Addressing the harm requires changes across multiple layers. Biases can become systemic, making short-term fixes insufficient, and the complexity may paralyze efforts for recourse, repair, \& prevention. 

Though more extreme cases may arise, seemingly minor upstream changes—like the one described---can trigger disproportionately large and harmful downstream effects; faulty software updates from cybersecurity company CrowdStrike, for example, caused millions of travelers to be grounded in July 2024, leading both SouthWest and Delta to sue CrowdStrike for hundreds of million in damages---despite existing best practices in software development and digital supply chains. Yet best practices \textit{do not exist} for AISCs. So, what forms of redress are there for the sepsis model provider, the hospitals, and customers? 

As mentioned in Section \ref{redress}, redress occurs when it is \textit{achievable} and there is \textit{consensus} between involved parties. The extent of the redress depends on the significance of the harm and the power of the affected stakeholders. In an ideal setting, not only is the harm stopped (recourse) and fixed (repair), and damages such as pain and lost income are compensated for (reparation), and the harm is learned from to prevent similar events in the future (prevention). Below, we detail the roles and incentives of each stakeholder group participating in the AISC before discussing how the AISC's structure affects stakeholders' ability to implement redress.


\paragraph{(A) EHR Platform.} As both an intermediary and a data provider, the EHR platform interfaces directly with each stakeholder. Their priority is to ensure that there are no failures on the part of their system, and that they are not liable for the failures of the other stakeholders. If the EHR organization is not integrated with other levels of the AISC, their participation in redress is as a potential mediator or in collaboration with other stakeholders.
As the EHR system interfaces with all stakeholders, implementing AISC-wide forms of traceability such as disclosures or  benchmarking to support minimally harmful updates requires their buy-in. 

\paragraph{(B) Voice-to-text model provider.} It is possible that B is not aware of their impact on the sepsis model provider (C). They are incentivized to improve their own product offerings, while minimizing avenues of redress that could lead to losses or the sharing of proprietary information. If there is sufficient traceability across the AISC, they may be held liable for redressments. While they may support collaborative efforts to prevent future negative outcomes, other forms of redress are antithetical to their interests.

\paragraph{(C) Sepsis model provider.} The sepsis model provider may lose the FDA medical clearance that allows them to operate, may lose customers, and could be sued by patients or hospitals. They desire a functional product and to reduce costs associated with their model's poor performance, yet they may not be aware of the existence of an upstream stakeholder, or that their product was impacted by an upstream change. If C is able to trace what caused their model's dysfunction, there are several actions they might take. First, they may opt for recourse, stopping the effects of the upstream voice-to-text model by no longer using dictated doctor notes or requesting that its provider reverts to an older version of the model. Alternatively, 
they may seek repairs in their upstream dependencies (e.g., voice-to-text performance is improved through an update) or by ensuring their model is robust to upstream changes. They may look for reparations given income or reputational losses. 
Finally, preventing re-occurrences might involve evaluation or disclosures implemented across AISC deployed updates. 

\paragraph{(D) Hospitals \& Patients.}
Hospitals and patients desire functioning products that do not lead to poor healthcare. To this end, hospitals may stop a harm from continuing (recourse) by discontinuing model use. This might be for sepsis and/or voice-to-text models, at least until action is taken to repair the undesired behavior. Hospitals and patients may seek reparations for damages (reputational, fiscal, physiological, etc.). This may be pursued through provider-customer problem resolution (for hospitals) such as licensing discounts or through legal action to receive monetary compensation (hospitals and patients).
Who provides reparation, and if reparation is even possible, depends on \textit{if the harm is traceable} (and thus the responsible party is clear), the dynamics of the AISC, and the urgency of the harm. Preventing future similar harms from happening may be difficult if the stakeholders are either not aware of each other or will not collaborate, though the hospital may simply adopt policies or request tooling to reduce reliance on the models.

\vspace{2mm}
\noindent The AISC's structure affects whether the response is materially achievabile. Similarly, market dynamics shapping the AI supply chain will influence the consensus determining if a response will be taken. Below, we discuss how markets yield different transparencies, costs, and bargaining positions, which in turn condition the available responses.

\subsection{Vertical Integration}
 Consider the AISC with vertical integration (Figure \ref{fig:marketsss} A). While there are other vertically integrated competitors, interoperability between components does not non-existent
 \footnote{Note that this form of vertical integration is canon to the AISC; model providers, infrastructure providers (both in cloud services and chip manufacturing), and data-cum-model providers such as social media companies alike have initiated integrations for all other stakeholders (except users of course). For interested readers, a compilation of vertically integrated examples is included in the Appendix.}. 
 This form of integration has several consequences. First, competing with a single product is difficult. As a result, competitors will also have vertically integrated offerings, reducing optionality for hospitals to select what AI products and services best suit their performance requirements. While this outcome aligns with established economic theory, its implications for AISCs are distinct. When multiple systems are tightly interconnected, identifying and addressing specific errors becomes increasingly complex. This complexity is magnified in data/model pipelines, and further exacerbated by systems like multimodal models \cite{dang2024explainable}. 
 Homogenization in the integrated ecosystem limits the variety of approaches available to mitigate robustness issues. A single monolith may rely on a unified architecture, which, while efficient, makes the system more vulnerable to systematic errors.
 These models add layers of opacity to already complex systems, making explanations that can be mapped causally across modalities difficult. Vertical integration, with its resources and unified interests, can encourage increased cross-modality AI. 

Second, redress is dominated by the priorities of the integrated provider. As vertical integration limits the ability of non-integrated stakeholders (e.g., hospitals) to reject individual features within a bundled offering, recourse is constrained. Switching providers is less viable due to the prohibitive cost, complexity, and disruption such a transition entails. There is one monolith aware of each participating product in the AISC, suggesting repairs may be more tractable due to the provider’s control over the entire ecosystem. Rather than require isolating failures, for example, the integrated firm can instead rely on large scale processes for deployment processes, including A/B testing. 
Providers, shielded by their control of the ecosystem, have little incentive to admit fault or offer compensation. In our scenario, a voice-to-text model misinterprets clinical inputs leading a downstream sepsis model to make inaccurate predictions. When a harm is incurred, the provider may more easily deflect blame or argue that the issue lies with the hospital's implementation rather than the product itself. Prioritizing internal efficiency over explanations or externally accountable mechanisms creates barriers for hospitals and patients seeking reparations. Further, repairs are driven by the provider’s business goals rather than the needs of hospitals or patients. Misalignments between the integrated firm and its customers may leads to delays, incomplete solutions, or fixes aimed at a broader customer base rather than addressing localized harm. Unless the harm is sufficiently urgent---patient health outcomes widely worse, or regulatory intervention is being considered---it will be deprioritized. Because decisions are unilaterally made by the integrated provider and are hidden by vertical walls, other actors have little power and fewer avenues for redress.



\subsection{Horizontal Integration}
In a horizontally integrated AISC (Figure \ref{fig:marketsss} B), the sepsis, voice-to-text, and other upstream models are controlled by a single provider, while EHR systems and hospitals remain independent. In vertically integrated systems, managing updates and resolving dependencies is relatively straightforward---limited only by technical constraints---but external transparency is absent unless mandated by regulation. In contrast, horizontal integration introduces an intermediary layer: the EHR system, which can observe changes from individual ML models. This position may aid in identifying harmful changes and negotiating repairs. Limited competition can reduce incentives for providers to offer meaningful redress, however, especially if hospitals lack alternatives. For example, providers may bundle services, preventing hospitals from opting out of individual models. Thus, recourse may still require abandoning the entire bundle rather than just the faulty component. 

A key difference between horizontal and vertical markets lies in the implementation of preventative strategies. In vertically integrated systems, the entity developing AI models also owns or directly manages the downstream systems in which those models are deployed. This structural alignment simplifies coordination: as the same organization bears the cost of both failures and fixes, preventative strategies---such as version control, staged rollouts, or pre-deployment testing---can be implemented unilaterally and prioritized according to internal objectives. However, this arrangement may not align with the needs of external stakeholders such as hospitals or patients, and it typically lacks external transparency unless mandated by regulation.

In contrast, horizontally integrated systems distribute responsibility across organizational boundaries: model developers, EHR vendors, and hospitals each control a different segment of the AISC pipeline. This fragmentation can introduce friction in adjudicating competing priorities---particularly when safety, usability, and liability span across entities with unequal power and incentives. Yet, if there is an intermediary layer---the EHR vendor, in this example---this integration also offers a unique advantage: greater visibility into both upstream model behavior and downstream clinical impact. This positioning creates new opportunities for transparency and monitoring that are often unavailable in vertically integrated systems. While the EHR may lack direct control over models, it may instrument tools to detect harmful changes early and facilitate redress, provided that the necessary incentives, infrastructure, and cooperation are in place.

\subsection{Free Market}
In a free market, external transparency and traceability are determined by buy-in from stakeholders, their IP concerns, and the technical expressiveness or reliability of model outputs (e.g., predictions or generated content). Unlike in horizontal and vertical integration, which can implement internal systems aligned with business priorities, free market participants may struggle with avenues of redress that work, as the lack of integration can reduce transparency between roles in the AISC. In our example, the hospital decides to individually contract with each model provider. If they are dissatisfied with the outcomes of the sepsis model, they can achieve recourse by switching to another. This will not change the performance of the voice-to-text model, however, and if the new sepsis model is dependent on the same unchanged voice-to-text model, the switch may not help. For this reason, repairs may be difficult without AISC transparency. If explanations are viable and all actors are known, reparation through litigation is relatively straightforward: actors know who is involved, and explanations are available to attribute liability. When the market consists of independent, non-integrated firms, and few standards to enforce transparency, actors in the AISC \textit{might not be aware of each other}. 

Resulting information asymmetries across AISCs can enable integration, particularly under unprotected free market settings \cite{akerlof1970lemons, north1990institutions}. For example, details about AISC configurations may be known to one stakeholder but not to others. This may enable adversarial actions---the voice-to-text provider may  implement decisions or learn features that negatively affect a downstream developer's product safety or reliability, making it easier for them to ensure specific partners are downstream, or allowing them to build competitive products of their own. Still, free markets can encourage interoperability and prevention strategies against failures. Cell phone providers adopted shared standards for sending and receiving packets (information) in part because there was no global monopoly. Interoperability reduces frictions for all recourse actions, as stakeholders are heavily incentivized to satisfy their customer needs and ease customers' transitions into their ecosystem. 
Similarly, reparations discounts or refunds are likely under open competition.
 As a result, buy-in for implementing best practies is high because there are numerous smaller firms benefiting from the standardization and clear liability.

\section{Discussion}
There are numerous actors participating in AI production and use. As technology is adopted, it produces both enhancements and errors to ongoing production systems---material or service. In designing functional AI supply chains, we must consider the factors challenging their resiliency. Our efforts orient towards four fundamental components of business since the late 19th century: competition, responsibility, liability, and transparency. This paper contributes not only theoretical insight but also actionable language and conceptual frameworks that are critical for designing more resilient AI supply chains. By introducing a clear typology of redress (recourse, repair, reparation, prevention) and mapping it against hypothetical market structures, we provide practitioners and policymakers with tools to diagnose vulnerabilities and opportunities for productive, responsible practices within AISCs. 

Redress is not simply a procedural fix but a basic communicative challenge, requiring complex, multi-faceted, and often dynamic transactions. This paper attempts to clarify how different stakeholders have varying capacities for offering or demanding redress, providing a foundation for targeted governance and design efforts. We emphasize agency, dialogue, and transparency from the perspective of both the impacted subject and the responsible party, interrogating how discourse is structured or constrained---what kinds of appeals are intelligible, to whom, and under what power asymmetries. As we show, both those harmed and those causing harm may be limited by their position within an AISC and by prevailing market power structures. Such constraints shape the ability to seek redress, yes, but also the ability to offer remedies to poor practices.

While the challenges facing AI supply chains seem unprecedented, they can nonetheless be effectively simulated by leveraging existing experience from rules-based monetary systems and consumer product supply chains. This prompts the question: What do we already know that can inform alternative approaches? Some observers suggest fostering competition through subsidies for small businesses, open-source models and datasets, and subsidized data marketplaces with strict terms of use.  Evidence from supply chain management and rules-based monetary systems provides a basis for simulating consequences of conditional variations on prior regulatory structures. 
The AI landscape is not monolithic; it comprises a heterogeneous mix of vertically integrated firms, horizontally organized intermediaries, and independent stakeholders. Designing resilient infrastructure for AI deployment must reflect this diversity.
Our stakeholder analysis illustrates a small fraction of this complexity--and hints at possible modes of intervention. 
 Lessons from regulated industries, such as the financial sector, suggest that hybrid models--where responsibility is shared and enforced through centralized licensing or procurement requirements--may offer a scalable path forward. However, the consequences of such interventions, including cease-and-desist orders, injunctions, and punitive damages, are not evenly distributed across stakeholders. Effective governance must balance short-term accountability with long-term system resilience, working to prevent structural bias and suppress homogenization while preserving technical experimentation and equitable access to innovation.
As the AI industry matures, it is crucial to design controls that balance transparency, traceability, and accountability while fostering innovation and competition.

\vspace{2mm}
\noindent \textbf{Limitations.}
Stakeholder theory has a complex relationship with concerns of distributive justice, and often faces empirical difficulties when establishing who is a stakeholder. Thus, stakeholder analysis often under-theorizes topics like resistance, marginality, or contestation, often assuming consensus is ideal. Although this is often a limitation, through our case studies we attempt to mitigate this by illustrating how stakeholder positionality is informed by structural asymmetries and institutional constraints.
Finally, while the paper draws on organizational, product, and service data from real companies, it does not include formal empirical methods such as interviews, surveys, or fieldwork. As a result, the stakeholder roles and redress typology are based on conceptual synthesis from existing published scholarship rather than through stakeholder accounts or case-based evidence.

\begin{acks}
We gratefully acknowledge Eppa Rixey, Jillian Ross, Luis Videgaray, Andrew Ilyas, Sarah Cen and Aleksander Madry for thoughtful discussions on AI supply chains. This work was supported by the MIT Social and Ethical Responsibilities of Computing (SERC) 2024 Seed Grant. 
\end{acks}

\bibliographystyle{ACM-Reference-Format}
\bibliography{references}


\appendix
\onecolumn

\renewcommand{\arraystretch}{1.2}
\begin{table}[H]
\caption{How each market dynamic affects the achievability and relational power around every route to redress for the harm of increased adverse health outcomes incurred by (D). Urgency of the harm is constant in all settings.}
\label{aisc_redress_health}
\centering
\begin{tabular}{|p{1.8cm}|p{3.9cm}|p{3.9cm}|p{3.9cm}|p{2.9cm}|}
\hline
\textbf{Market} & \textbf{Recourse} & \textbf{Repair} & \textbf{Reparation} & \textbf{Prevention} \\
\hline

\textbf{Horizontal} &
\parbox[t]{3.9cm}{
\colorbox{green}{Achievable}: Yes. Stop using (C).\\
\colorbox{orange}{Urgency}: How many adverse outcomes? How adverse?\\
\colorbox{yellow}{Power}: Reduced optionality prevents opting-out of features. Must end use of (C) \textit{and} (B).
} &

\parbox[t]{3.9cm}{
\colorbox{green}{Achievable}: Yes. Even if mechanism of bias is unrecoverable, traceable through A/B testing.\\
\colorbox{yellow}{Power}: High bar for activation from model provider. Depends on regulation or collective experiences from more customers.
} &

\parbox[t]{3.9cm}{
\colorbox{green}{Achievable}: Yes, through litigation. Liability is provable.\\
\colorbox{yellow}{Power}: (D) has low relative power. Difficult to negotiate beneficial terms in licensing agreement, therefore harm must be egregious.\\
\colorbox{green}{Achievable}: Yes, through discounts or refunds.\\
\colorbox{yellow}{Power}: (D) has low relative power (as one of many customers) to demand such reparations.
} &

\parbox[t]{2.9cm}{
\colorbox{green}{Achievable}: Yes. Reduce reliance on (C) for septic diagnostics.\\
\colorbox{yellow}{Power}: Single-party decision.
} \\
\hline

\textbf{Vertical} &
\parbox[t]{3.9cm}{
\colorbox{green}{Achievable}: Yes. See above.\\
\colorbox{yellow}{Power}: Reduced optionality prevents opting-out of features. May be unable to stop paying for (C) or (D).\\
\colorbox{green}{Achievable}: Yes. Switch from (A) to (A')\\
\colorbox{yellow}{Power}: High switching costs from ecosystem.
} &

\parbox[t]{3.9cm}{
\colorbox{green}{Achievable}: See above\\
\colorbox{yellow}{Power}: See above. More relative power to (D) due to other competing vertically integrated firms that collectively set standards and update norms.
} &

\parbox[t]{3.9cm}{
\colorbox{green}{Achievable}: Yes, through litigation. Liability is provable.\\
\colorbox{yellow}{Power}: (D) has less relative power to demand insight into integrated system, and may not have enough information. However, it is easier to compare to other firm norms to prove liability.\\
\colorbox{green}{Achievable}: Yes, through discounts or refunds.\\
\colorbox{yellow}{Power}: Greater competition gives (D) more relative power than in horizontal integration.
} &

\parbox[t]{2.9cm}{
\colorbox{green}{Achievable}: Yes. See above.\\
\colorbox{yellow}{Power}: See above.
} \\
\hline

\textbf{Free Market} &
\parbox[t]{3.9cm}{
\colorbox{green}{Achievable}: Yes. Switch from (C) to (C')\\
\colorbox{yellow}{Power}: Single-party decision.\\
\textit{Note: Though this stops the current harm, (D) might experience the same issues if (C') unknowingly also uses (B).}
} &

\parbox[t]{3.9cm}{
\colorbox{green}{Achievable}: Maybe, if (C) can be fixed w/out changes to voice to text. Otherwise, A/B testing is impossible and mechanism of bias is unrecoverable.\\
\colorbox{yellow}{Power}: (D) has high relative power due to increased competition and lower switching costs.
} &

\parbox[t]{3.9cm}{
\colorbox{green}{Achievable}: Maybe, if supply chain transparency is sufficiently great or issue is sufficiently shallow such that liability can be proven.\\
\colorbox{yellow}{Power}: (D) has relatively high power to demand beneficial contracts that enable litigation.\\
\colorbox{green}{Achievable}: Yes, through discounts or refunds.\\
\colorbox{yellow}{Power}: High competition gives (D) significant relative power to demand discounts or refunds due to low switching costs.
} &

\parbox[t]{2.9cm}{
\colorbox{green}{Achievable}: See above.\\
\colorbox{yellow}{Power}: See above.
} \\
\hline
\end{tabular}
\end{table}

\renewcommand{\arraystretch}{1.2}
\begin{table}[H]
\caption{How market dynamics affect every route to redress for the harm of fiscal losses incurred by (C). Urgency of the harm is constant for all responses.}
\label{markets_sepsis}
\centering
\begin{tabular}{|p{1.8cm}|p{3.9cm}|p{3.9cm}|p{2.9cm}|p{3.9cm}|}
\hline
\textbf{Market} & \textbf{Recourse} & \textbf{Repair} & \textbf{Reparation} & \textbf{Prevention} \\
\hline

\textbf{Horizontal} &
\parbox[t]{3.9cm}{
\colorbox{green}{Achievable}: Yes. Revert back to previous version of model.\\
\colorbox{orange}{Urgency}: Relatively low where one or few customers’ dissatisfaction is unimpactful.\\
\colorbox{yellow}{Power}: Single-party decision.
} &

\parbox[t]{3.9cm}{
\colorbox{green}{Achievable}: Yes. Fix (C) to work with (B) through A/B testing if untraceable.\\
\colorbox{yellow}{Power}: Single-party decision.
} &

\parbox[t]{2.9cm}{
Irrelevant
} &

\parbox[t]{3.9cm}{
\colorbox{green}{Achievable}: Ensure greater model behavior robustness or flag upstream changes for possible downstream impacts.\\
\colorbox{yellow}{Power}: Requires buy-in from (A). (C) has reasonable relational power.
} \\
\hline

\textbf{Vertical} &
\parbox[t]{3.9cm}{
\colorbox{green}{Achievable}: Maybe. Stop using features from (B) if their impact is traceable.\\
\colorbox{orange}{Urgency}: Relatively low where one or few customers’ dissatisfaction is unimpactful.\\
\colorbox{yellow}{Power}: Single-party decision.
} &

\parbox[t]{3.9cm}{
\colorbox{green}{Achievable}: Yes. Fix (C) to work with (B) through A/B testing if untraceable.\\
\colorbox{yellow}{Power}: Single-party decision.
} &

\parbox[t]{2.9cm}{
Irrelevant
} &

\parbox[t]{3.9cm}{
\colorbox{green}{Achievable}: Ensure greater model behavior robustness or flag upstream changes for possible downstream impacts.\\
\colorbox{yellow}{Power}: Single-party decision.
} \\
\hline

\textbf{Free Market} &
\parbox[t]{3.9cm}{
\colorbox{green}{Achievable}: Yes. Revert back to previous version of model.\\
\colorbox{orange}{Urgency}: Higher due to increased competition.\\
\colorbox{yellow}{Power}: Single-party decision.
} &

\parbox[t]{3.9cm}{
\colorbox{green}{Achievable}: Yes. Fix (C) to work with (B) through A/B testing if untraceable.\\
\colorbox{yellow}{Power}: Single-party decision.
} &

\parbox[t]{2.9cm}{
Irrelevant
} &

\parbox[t]{3.9cm}{
\colorbox{green}{Achievable}: Ensure greater model behavior robustness or flag upstream changes for possible downstream impacts.\\
\colorbox{yellow}{Power}: Buy-in needed. (C) has greater relational power.
} \\
\hline
\end{tabular}
\end{table}

\begin{table}[h!]
\centering
\caption{Market Dynamics and Associated Harms}
\begin{tabular}{|>{\centering\arraybackslash}m{3cm}|>{\centering\arraybackslash}m{3cm}|>{\centering\arraybackslash}m{3cm}|>{\centering\arraybackslash}m{3cm}|>{\centering\arraybackslash}m{3cm}|}
\hline
\textbf{Market} & \textbf{Biased Decision Making} & \textbf{Reduced Optionality} & \textbf{Homogenization}  & \textbf{Poor Explanations or Traceability}  \\ \hline
Horizontal & \checkmark & \checkmark & \checkmark & \checkmark \\ \hline
Vertical & \checkmark & \checkmark & & \checkmark \\ \hline
Free Market & \checkmark & & & \checkmark \\ \hline
\end{tabular}
\label{harmsinaisc}
\end{table}

\section{Additional Characterization of AISC Stakeholders}
\label{sec:extendedstakeholders}

Here, we provide an extended characterization of roles performed by stakeholders participating within AI supply chains (AISCs). Our categorization synthesizes existing datasets, prior literature, and domain knowledge, delineating stakeholder roles by their contributions to AI production. As described previously, we focus on characterizing endogenous actors, or those directly engaged within the AI supply chain. We do not characterize exogenous actors, or those operating from outside a functioning supply chain (e.g. legislatures, certain researchers, or auditors). Stakeholder roles described here are analytical categories reflecting functions rather than exclusive organizational identities. Entities frequently engage in multiple roles simultaneously, contributing to dynamic interdependencies and complexities within AI supply chains. 
Below, we provide a detailed characterization of each endogenous stakeholder type.

\subsection{Infrastructure Providers}

\begin{figure}[h]
  \raggedright
\includegraphics[width=.5\linewidth]{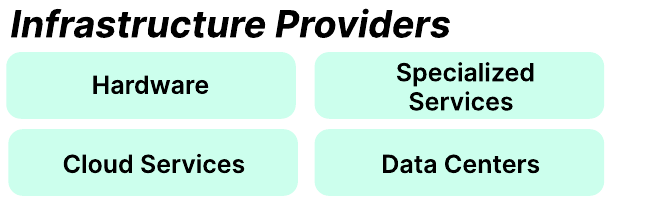}
    \Description{A white‐background box titled “Infrastructure Providers” containing four light-green rounded rectangles arranged in two rows: 
    Top left reads “Hardware”; top right reads “Specialized Services”; bottom left reads “Cloud Services”; bottom right reads “Data Centers.”}
    \label{fig:infrastructure}
    \vspace{-3mm}
\end{figure}

Infrastructure providers supply fundamental technological tools and resources required for AI development, deployment, and management.

\subsubsection*{Hardware Providers}
Companies like Intel, ARM, IBM, Qualcomm, and Nvidia manufacture physical computing components essential for model training and deployment, including CPUs, GPUs, TPUs, and integrated circuits.

\subsubsection*{Data Centers and Server Farms}
These facilities house extensive clusters of interconnected computers and storage units necessary for large-scale computational tasks. Access is often monopolized by large corporations due to high leasing costs, limiting smaller entities’ independence and flexibility.

\subsubsection*{Cloud Services}
Firms such as Amazon Web Services (AWS), Microsoft Azure, and Google Cloud Platform provide scalable computational resources, including compute instances, storage, networking, and specialized AI tools. Their integrated services foster lock-in and dependency among downstream users.

\subsubsection*{Specialized Services}
Emerging services like edge computing, IoT infrastructure, and federated learning cater to niche needs for privacy, security, and efficiency, intersecting frequently with traditional infrastructure providers.

\subsection{Data Providers}
\begin{figure}[h]
  \raggedright
\includegraphics[width=.5\linewidth]{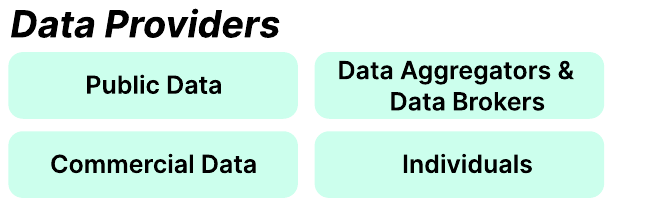}
    \Description{A white‐background box titled “Data Providers” containing four light-green rounded rectangles in a two-by-two grid: 
    Top left “Public Data,” top right “Data Aggregators \& Data Brokers,” bottom left “Commercial Data,” bottom right “Individuals.”}
    \label{fig:data}
    \vspace{-3mm}
\end{figure}

Data providers offer critical datasets, enabling model training, fine-tuning, and analytics.

\subsubsection*{Commercial Data Providers}
Firms like Experian aggregate proprietary datasets covering consumer behavior, market research, and finance, monetizing access through structured contracts and data marketplaces. Data marketplaces facilitate transactions between sellers and buyers, each governed by unique licensing terms. Specialized personal data marketplaces have emerged to empower individuals economically and mitigate privacy infringements historically associated with mass data collection.

\subsubsection*{Public Data Providers}
Organizations providing publicly accessible datasets (e.g., CommonCrawl, government census data) support transparency and democratic information access. Data collection mechanisms include sensor-based collection, web scraping, and first-party platform data collection, with ongoing legal and ethical debates about ownership and usage rights.

\subsection{Model Providers}
\begin{figure}[h]
\raggedright
\includegraphics[width=.6\linewidth]{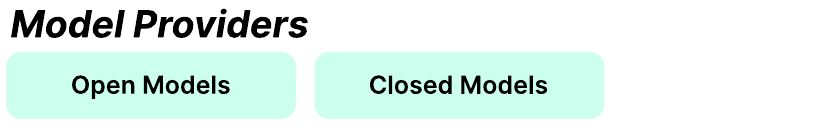}
    \Description{A white‐background box titled “Model Providers” containing two light-green rounded rectangles side by side: left reads “Open Models,” right reads “Closed Models.”}
    \label{fig:models}
    \vspace{-3mm}
\end{figure}

Model providers develop AI models, varying widely in openness and accessibility. Providers range from non-profit consortiums (EleutherAI, BigScience) to commercial entities (OpenAI, Meta). Despite the proliferation of providers, financial and market power remain concentrated among major corporations. Their dominance concentrates market and user attention, affecting supply chain dynamics through their distribution and access policies.

\subsubsection*{Open Models}
Publicly accessible models with permissive licensing (e.g., EleutherAI’s GPT-J, BigScience’s BLOOM) support broad developer collaboration, but may lack long-term support.

\subsubsection*{Proprietary Models}
Controlled models hosted by corporations like OpenAI and Meta, typically accessed via restricted APIs or exclusive partnerships, significantly shape market dynamics through platform lock-in.

\subsection{Intermediaries}
\begin{figure}[h]
\raggedright
\includegraphics[width=.9\linewidth]{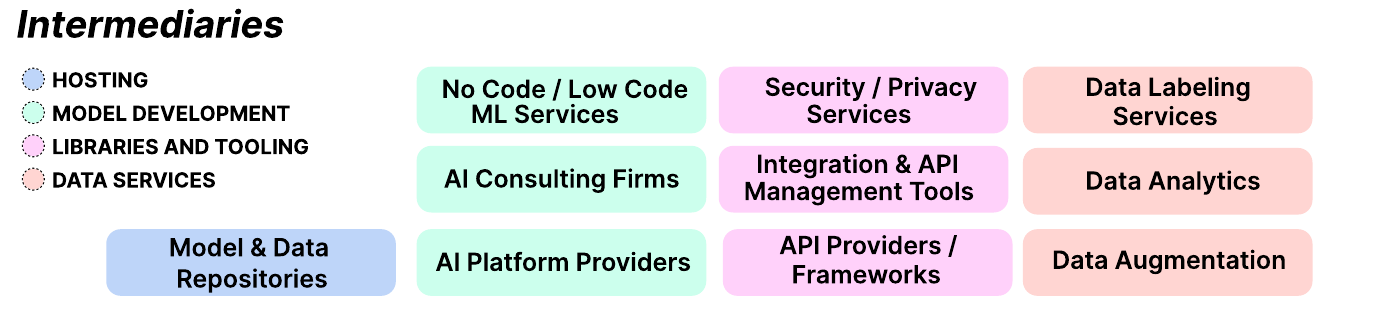}
    \Description{A white‐background panel titled “Intermediaries.” On the left is a legend of four dotted‐circle icons labeled “Hosting,” “Model Development,” “Libraries and Tooling,” and “Data Services.” To the right is a tiled layout of colored rounded rectangles, grouped by category:
    \begin{itemize}
      \item Light green: “No Code / Low Code ML Services,” “AI Consulting Firms,” “AI Platform Providers.”
      \item Light purple: “Security / Privacy Services,” “Integration \& API Management Tools,” “API Providers / Frameworks.”
      \item Light red/pink: “Data Labeling Services,” “Data Analytics,” “Data Augmentation.”
      \item Light blue: “Model \& Data Repositories.”
    \end{itemize}}
    \label{fig:intermediaries}
    \vspace{-3mm}
\end{figure}

Intermediaries offer specialized, non-user-facing services within AISCs, increasing supply chain complexity.

\subsubsection*{Data Services}
Include data labeling (e.g., Scale, Amazon Turk), synthetic data generation, and augmentation to enhance dataset quality or diversity.

\subsubsection*{Model Development and AI Services}
Provide low-code/no-code ML services (e.g., DataRobot, Runway), consultancy, and deployment platforms (e.g., Google AutoML, Amazon SageMaker), facilitating rapid and accessible AI product development.

\subsubsection*{Model Hosting}
Repositories such as Hugging Face and TensorFlow Hub offer centralized access to large-scale models, easing storage, discovery, and integration with cloud services.

\subsubsection*{Libraries and Tooling}
Essential open-source frameworks and libraries (e.g., PyTorch, TensorFlow, Scikit-learn) offer foundational tools for AI development. Major firms support these projects for strategic market influence, simultaneously benefiting from community-driven improvements.

\subsection{User/Consumer-Facing Products}

\begin{figure}[h]
  \raggedright
\includegraphics[width=.7\linewidth]{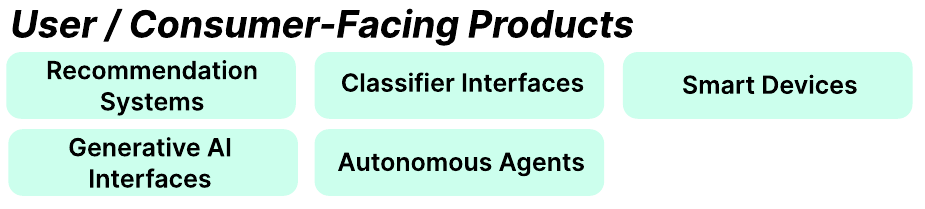}
    \Description{A white‐background box titled “User / Consumer-Facing Products” containing five light-green rounded rectangles arranged in two rows: 
    Top row “Recommendation Systems,” “Classifier Interfaces,” “Smart Devices.” 
    Bottom row “Generative AI Interfaces,” “Autonomous Agents.”}
    \label{fig:products}
    \vspace{-3mm}
\end{figure}

User-facing AI applications directly impact user experiences, frequently shaping user interactions and decisions. Broadly, user-facing artifacts may include open-software as well as user-facing products available only through purchase.  We do not distinguish consumer- and user-facing products in this and the following stakeholder category, except when the distinction is necessary for a particular point. User- and consumer- facing products can be differentiated by their associated interfaces in which they are embedded or mediated.

\subsubsection*{Recommendation Systems:} Predominantly utilized by social media, search engines, and e-commerce platforms (e.g., Amazon, Shein), these systems influence consumer behavior by algorithmically curating content, often leading to homogenization and limiting user autonomy.
\subsubsection*{Classifier Interfaces:} Widely deployed for security, anomaly detection, and identity verification (e.g., Clear, TSA), classifiers mediate user interactions, frequently through expert oversight, thus indirectly affecting end users.
\subsubsection*{Smart Devices:} IoT devices embedded in daily life (e.g., smartwatches, home systems) automate monitoring and data collection, facilitating substantial user-data generation. Applications span personal convenience to infrastructure maintenance and urban planning, with notable privacy implications.
\subsubsection*{Generative AI Interfaces:} Tools built upon LLMs and diffusion models enable user-driven content generation (e.g., virtual try-ons, image editing). Although powerful, these tools also amplify privacy concerns and potential misuse.
\subsubsection*{Autonomous Agents:} These digital and physical systems (e.g., autonomous vehicles, digital assistants) independently execute tasks. Their growing complexity and autonomy raise new concerns about accountability, liability, and personhood, highlighting regulatory and ethical challenges.

\subsection{Users \& Consumers}
\begin{figure}[h]
  \raggedright
\includegraphics[width=.6\linewidth]{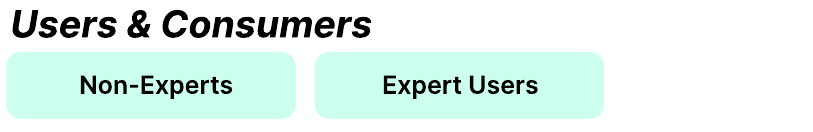}
    \Description{A white‐background box titled “Users \& Consumers” containing two light-green rounded rectangles side by side: “Non-Experts” on the left and “Expert Users” on the right.}
    \label{fig:users}
    \vspace{-3mm}
\end{figure}

Users constitute the largest and most diverse stakeholder group, shaping AISCs through varied interactions.

\subsubsection*{Non-Expert Users}
Frequently unaware of underlying AI operations, non-expert users primarily interact through default settings, often influenced by design choices (e.g., dark patterns). This limited awareness can inhibit informed decisions and complicate opting-out processes.

\subsubsection*{Expert Users}
Possessing specialized knowledge, expert users critically influence system adoption and refinement. Their domain expertise empowers them to navigate, adapt, or reject AI products, influencing providers to meet specialized, informed demands. Expert feedback frequently drives innovation and customization within niche applications.

\section{Existing Examples of Vertical Integration}
\label{sec:vertical}

The following cloud service providers offer services catering to various business needs, from basic infrastructure and storage to advanced AI, machine learning, and big data solutions. This is only a sampling of vertically integrated firms.

\subsubsection{Amazon Web Services (AWS)} \hfill\\
\textbf{Compute:} Amazon EC2 (Elastic Compute Cloud), AWS Lambda (serverless computing)\\
\textbf{Storage:} Amazon S3 (Simple Storage Service), Amazon EBS (Elastic Block Store), Amazon Glacier (archival storage)\\
\textbf{Databases:} Amazon RDS (Relational Database Service), Amazon DynamoDB (NoSQL), Amazon Aurora\\
\textbf{Networking:} Amazon VPC (Virtual Private Cloud), Amazon Route 53 (DNS service), AWS Direct Connect\\
\textbf{AI/ML:} Amazon SageMaker, AWS Deep Learning AMIs\\
\textbf{Big Data:} Amazon EMR (Elastic MapReduce), Amazon Redshift (data warehousing)\\
\textbf{IoT:} AWS IoT Core, AWS Greengrass\\
\textbf{Security:} AWS Identity and Access Management (IAM), AWS Key Management Service (KMS)\\
\textbf{Developer Tools:} AWS CodePipeline, AWS CodeBuild, AWS CodeDeploy

\subsubsection{Microsoft Azure} \hfill\\
\textbf{Compute:} Azure Virtual Machines, Azure Functions (serverless), Azure Kubernetes Service (AKS)\\
\textbf{Storage:} Azure Blob Storage, Azure Disk Storage, Azure Archive Storage\\
\textbf{Databases:} Azure SQL Database, Azure Cosmos DB (multi-model), Azure Database for PostgreSQL/MySQL\\
\textbf{Networking:} Azure Virtual Network, Azure DNS, Azure ExpressRoute\\
\textbf{AI/ML:} Azure Machine Learning, Cognitive Services (e.g., vision, speech, language APIs)\\
\textbf{Big Data:} Azure Synapse Analytics, Azure HDInsight\\
\textbf{IoT:} Azure IoT Hub, Azure Sphere\\
\textbf{Security:} Azure Active Directory, Azure Security Center\\
\textbf{Developer Tools:} Azure DevOps, Azure DevTest Labs

\subsubsection{Google Cloud Platform (GCP)} \hfill\\
\textbf{Compute:} Google Compute Engine, Google Cloud Functions, Google Kubernetes Engine (GKE)\\
\textbf{Storage:} Google Cloud Storage, Persistent Disks, Google Cloud Filestore\\
\textbf{Databases:} Google Cloud SQL, Google Cloud Spanner, Google Bigtable (NoSQL)\\
\textbf{Networking:} Google Virtual Private Cloud (VPC), Cloud DNS, Cloud Interconnect\\
\textbf{AI/ML:} AI Platform, TensorFlow, AutoML\\
\textbf{Big Data:} BigQuery, Dataflow, Dataproc\\
\textbf{IoT:} Cloud IoT Core\\
\textbf{Security:} Identity and Access Management (IAM), Cloud Key Management Service (KMS)\\
\textbf{Developer Tools:} Google Cloud Build, Cloud Source Repositories, Cloud Deployment Manager

\subsubsection{IBM Cloud} \hfill\\
\textbf{Compute:} IBM Cloud Virtual Servers, Bare Metal Servers, IBM Cloud Functions (serverless)\\
\textbf{Storage:} IBM Cloud Object Storage, IBM Cloud Block Storage, IBM Cloud File Storage\\
\textbf{Databases:} IBM Db2, Cloudant (NoSQL), IBM Cloud Databases for PostgreSQL/MySQL\\
\textbf{Networking:} IBM Cloud Virtual Private Cloud, IBM Cloud Internet Services\\
\textbf{AI/ML:} IBM Watson, Watson Studio, Watson Machine Learning\\
\textbf{Big Data:} IBM Cloud SQL Query, IBM Cloud Analytics Engine\\
\textbf{IoT:} IBM Watson IoT Platform\\
\textbf{Security:} IBM Cloud Identity and Access Management, IBM Cloud Security Advisor\\
\textbf{Developer Tools:} IBM Cloud Continuous Delivery, Tekton Pipelines, IBM Cloud DevOps Insights

\subsubsection{Oracle Cloud} \hfill\\
\textbf{Compute:} Oracle Cloud Infrastructure (OCI) Compute, Oracle Functions (serverless)\\
\textbf{Storage:} Oracle Cloud Infrastructure Object Storage, Block Storage, Archive Storage\\
\textbf{Databases:} Oracle Autonomous Database, Oracle Database Cloud Service, MySQL Cloud Service\\
\textbf{Networking:} Oracle Virtual Cloud Network (VCN), Oracle Cloud Infrastructure FastConnect\\
\textbf{AI/ML:} Oracle AI, Oracle Machine Learning\\
\textbf{Big Data:} Oracle Big Data Service, Oracle Data Flow\\
\textbf{IoT:} Oracle IoT Cloud Service\\
\textbf{Security:} Oracle Identity and Access Management, Oracle Cloud Guard\\
\textbf{Developer Tools:} Oracle Developer Cloud Service, Oracle Visual Builder

\subsubsection{Alibaba Cloud} \hfill\\
\textbf{Compute:} Elastic Compute Service (ECS), Serverless Kubernetes, Function Compute\\
\textbf{Storage:} Object Storage Service (OSS), Elastic Block Storage, Archive Storage\\
\textbf{Databases:} ApsaraDB for RDS (MySQL/PostgreSQL/SQL Server), ApsaraDB for MongoDB, PolarDB\\
\textbf{Networking:} Virtual Private Cloud (VPC), Alibaba Cloud DNS, Express Connect\\
\textbf{AI/ML:} Machine Learning Platform for AI (PAI), Alibaba Cloud TensorFlow\\
\textbf{Big Data:} MaxCompute, DataWorks, AnalyticDB\\
\textbf{IoT:} Alibaba Cloud IoT Platform, Link Develop\\
\textbf{Security:} Anti-DDoS, Cloud Firewall, Identity and Access Management (IAM)\\
\textbf{Developer Tools:} Alibaba Cloud DevOps, CloudShell, Container Service for Kubernetes (ACK)
\noindent

\end{document}